\documentclass[nofootinbib,a4paper,aps,twocolumn,prd,10pt,superscriptaddress]{revtex4}

\usepackage{graphicx}
\usepackage{amsmath}
\usepackage{amsfonts}
\usepackage{amsthm}
\usepackage{mathrsfs}
\usepackage{amssymb}
\usepackage{epsfig}
\usepackage{amscd}
\usepackage{dcolumn}
\usepackage{bm}
\usepackage{natbib}
\usepackage{url}
\usepackage{xspace}
\usepackage{kantlipsum}

\usepackage[normalem]{ulem}
\usepackage{appendix}
\usepackage{algorithm}
\usepackage{algorithmic}

\usepackage{siunitx}
\usepackage{paralist}
\usepackage{placeins}

\usepackage{xcolor}

\usepackage{comment}

\usepackage{hyperref}
\hypersetup{%
    ,urlcolor=blue
    ,citecolor=blue
    ,linkcolor=blue
    }

\def\be{\begin{equation}}
\def\ee{\end{equation}}
\def\bea{\begin{eqnarray}}
\def\eea{\end{eqnarray}}

\def\prl{Phys. Rev. Lett.}

\def\prd{Phys. Rev. D}

\def\mnras{Mon. Not. Roy. Astr. Soc.}
\def\apj{Astrophys. J.}
\def\apjl{Astrophys. J. Lett.}

\def\aap{Astron. Astrophys.}
\def\actaa{Acta Astronomica}

\def\pasj{Publications of the Astronomical Society of Japan}

\def\physrep{Phys. Rep.}
\def\jcap{Journal of Cosmology and Astroparticle Physics}
\def\apss{Astrophysics and Space Science}

\begin{document}

\title{Numerical analysis of quasi-periodic oscillations with spherical spacetimes}

\author{Kuantay~Boshkayev}
\email{kuantay@mail.ru}
\affiliation{NNLOT, Al-Farabi Kazakh National University, Al-Farabi av. 71, 050040 Almaty, Kazakhstan.}
\affiliation{International University of Information Technology, Manas st. 34/1, 050040 Almaty, Kazakhstan.}
\author{Orlando~Luongo}
\email{orlando.luongo@unicam.it}
\affiliation{NNLOT, Al-Farabi Kazakh National University, Al-Farabi av. 71, 050040 Almaty, Kazakhstan.}
\affiliation{Scuola di Scienze e Tecnologie, Universit\`a di Camerino, Via Madonna delle Carceri 9, 62032 Camerino, Italy.}

\author{Marco Muccino} \email{muccino@lnf.infn.it}
\affiliation{NNLOT, Al-Farabi Kazakh National University, Al-Farabi av. 71, 050040 Almaty, Kazakhstan.}

\begin{abstract}
We numerically test quasi-periodic oscillations using three theoretically-motivated models
of spacetime adopting neutron star sources. Then, we compare our findings with a spherically-symmetric spacetime inferred from  $F(R)$ gravity, with constant curvature, showing that it fully-degenerates with our previous metrics, that have been adopted in the context of general relativity. To do so, we work out eight neutron stars in low mass X-ray binary systems and  consider a  Reisser-Nordstr\"{o}m solution plus a de Sitter phase with unspecified sign for the cosmological constant term. In particular, we investigate three hierarchies, \textit{i.e.}, the first dealing with a genuine Schwarzschild spacetime, the second with de Sitter phase whose sign is not fixed \emph{a priori} and, finally, a Reisser-Nordstr\"{o}m spacetime with an additional cosmological constant contribution. We perform Markov chain Monte Carlo analyses, based on the Metropolis-Hastings algorithm, and infer 1--$\sigma$ and 2--$\sigma$ error bars. For all the sources, we find suitable agreement with
spherical solutions with non-zero cosmological constant terms, \textit{i.e.}, with either de Sitter or anti-de Sitter solutions. From our findings, we notice that the existence of topological contribution to the net charge, suggested from $F(R)$ extensions of gravity, seems to be disfavored.  Finally, we focus on the physics of the cosmological constant term here involved, investigating physical consequences and proposing possible extensions to improve our overall treatments.
\end{abstract}

\maketitle

\section{Introduction}
\label{intro}

An increasing interest towards the physics of black holes (BHs) has recently been increased after the detection of gravitational wave \cite{2016PhRvL.116f1102A} and BH shadows \cite{2019ApJ...875L...1E}. The latter has put a step towards probing BHs in regimes where Einstein's theory may break down.
All the above outcomes have strengthened the idea that a possible epoch of \emph{BH precision astronomy} could start \cite{2021NatRP...3..732V}.

More broadly, by adopting compact objects, instead of BHs only, such as neutron star (NS),  white dwarf, etc. as possible sources of new data, one can find evidences about how stars evolve and configure, how gravity behaves in a strong field regime, and, more generally, hints about cosmological epochs in which the aforementioned objects have been formed \cite{2018ASSL..457....1C}. Indeed, joint gravitational and electromagnetic observations from NS merger have provided unprecedented insight into astrophysics, dense matter physics, gravitation and cosmology \cite{2017PhRvL.119p1101A}.

In addition to the above, low-mass X-ray binaries, where at least one compact object is a NS, as well as microquasars, exhibit analogous quasi-periodic oscillations (QPOs) in their X-ray flux \cite{Stella_1998, 1999PhRvL..82...17S,1999ApJ...524L..63S}, identified
as narrow peaks of excess energy in the corresponding
power spectra \cite{2004PASJ...56..553R, 2004ApJ...609L..63A, 2005A&A...436....1T,2012ApJ...760..138T,2014GrCo...20..233B}.
These peaks are associated with the process of matter accretion into the compact objects, therefore the investigation of QPOs can reveal the nature of the system under exam and/or direct NS properties, and the underlying gravity models \cite{2014AcA....64...45S, 2015ARep...59..441B, 2016ApJ...833..273T,2018mgm..conf.3433B, 2019MNRAS.488.3896T,2021Galax...9...75R}. Based on the frequency strength, QPOs can be classified into: 1) \emph{high frequency QPOs}, within the domain of $[0.1;1]$ kHz, and 2) \emph{low frequency QPOs}, showing frequencies smaller than $0.1$ kHz.

Several models of QPOs exists. Naively, at very high frequencies, the QPO characteristic frequencies appear close to the value of those of the test particle,
geodesic epicyclic oscillations around the gravitating compact object, suggesting that the scenario involving the
innermost stable circular orbit  of test particles
could be the most accredited to explain QPO nature\footnote{Very likely, the timescales of particle orbital motions near the compact object under exam provides the QPO frequencies.} \cite{2006csxs.book...39V,Lamb2008}. Consequently, the standard \emph{relativistic precession model}
suggests that QPOs arise from motion of
inhomogeneities due to accretion mechanism, say from blobs or vortexes \cite{1999PhRvL..82...17S,2012MNRAS.422.2581P}.

Motivated by the fact that general relativity (GR) in strong gravity regime can be tested by QPOs, we here consider a static spherical configuration provided by three physically-distinct cases. The first assumes a widely-consolidate Schwarzschild spacetime, namely the simplest approach that one can consider, the second its natural extension adding a unfixed sign of cosmological constant energy momentum term, namely the Schwarzschild--de Sitter or anti-de Sitter solutions, and lastly a (quasi) Reissner-Nordstr\"{o}m solution, where the charge is reinterpreted either as electric or topological. We then consider eight sources of NS in the low mass X-ray binaries from which we employ the most recent QPO data sets. We thus apply a Markov chain Monte Carlo (MCMC) analysis, deriving the best-fit parameters and the corresponding 1--$\sigma$ and 2--$\sigma$ errors, for each involved model, showing the most suitable ones capable of describing the aforementioned sources.
In particular, we notice that from extended fourth order gravity, it is possible to recover the form of the three involved spacetimes with a different physical meaning than the one inferred from GR. In extended scenarios, the charge is not given by the external electromagnetic field, but rather it is a topological charge. Even though the two frameworks fully-degenerate, considerations about the physics of our findings are discussed in both GR and extended theories of gravity. As a final conclusion, we notice that no charges are expected to be significant in the analyses, whereas the best treatments are those involving Schwarzschild-de Sitter (or anti-de Sitter) spacetimes. Further, evidence for no need of extending gravity, at this level, are also discussed and critically analyzed.

The paper is organized as follows. In Sec.~\ref{sezione2}, we describe QPOs in detail and how to compute them. The proposed spacetimes are reported and the physical motivations behind them are emphasized. In Sec.~\ref{sezione3}, the case of extended theories of gravity, in particular $F(R)$ paradigms, are discussed. We emphasize the degeneracy between our approaches in GR and the corresponding results obtained in that context. In Sec.~\ref{sezione4}, we perform MCMC analyses and discuss the physical interpretations of our findings. Further, we argue plausible consequences of our outcomes. Finally, Sec.~\ref{sezione5} shows up conclusions and perspectives of our work.

\section{Quasi-periodic oscillation}\label{sezione2}

Evidences for QPOs have been emphasized in the power spectra of the flux from X-ray binary
pulsars, and soon  after this first discovery, investigations in accretion disks were carried on \cite{2014PhRvD..89l7302B}. Consequently, QPOs point out how to test gravity and in general how to argue information from  compact objects and more broadly for cosmological purposes \cite{2009JCAP...09..013B}.
Indeed, current observational data got from accretion disks around compact
objects put tight frequency measures of QPOs, showing which model for astrophysical processes around compact gravitating objects is the most suitable one \cite{2011CQGra..28k4009M}.

In particular, QPOs depend on the background field equations involved through the use of the underlying spacetime. To clarify it better, assuming a given symmetry and a classical field theory that describes gravity, \textit{i.e.}, GR or alternatives to GR, it is possible to infer the spacetime that fulfills the symmetry itself. This spacetime will depend on the free parameters \emph{induced} by the field theory that we are employing. For instance, in the case of the Schwarzschild metric, the free term is a point-like mass $M$ only, describing the mass singularity or generally the compact object mass that we model through it. Thus, it appears evident that in order to test a gravitational theory by virtue of QPO data, one first has to fix the spacetime, getting it from prime principles applied to the alternative field equations  and second one can check how to fix the free parameters of the theory itself.

Motivated by the above considerations, we start with a generic spherically symmetric static spacetime composed of three main parameters: i) the point-like mass, $M$, ii) an effective vacuum energy term that modifies the energy momentum tensor by adding a constant contribution to the rest energy, iii) a possible charge, acting as the net charge contribution provided by the compact object.

We also reinterpret this metric by means of extended fourth order $F(R)$ theories of gravity, where this metric can represent a viable solution that fully degenerates with GR if one assumes a Reisser-Nordstr\"{o}m spacetime with de Sitter and/or anti-de Sitter non-vanishing term.

\subsection{Theoretical set up}

To get the
spherically-symmetric gravitational field providing both charge and vacuum energy we adopt a generic static spherically symmetric spacetime by \cite{Misner1973}
\begin{equation}
{\rm d}s^2 = -e^{\nu(r)}\,{\rm d}t^2+e^{\lambda(r)}\,{\rm d}r^2 + r^2({\rm
d}\theta^2+\sin^2\theta\,{\rm d}\phi^2)\,.
\end{equation}
Fixing $8\pi G =1$ and having the Einstein
equations under the form
\begin{subequations}
\begin{align}
{T_t}^t &= - e^{-\lambda}\left(\frac{1}{r^2} -
\frac{\lambda^\prime}{r}\right) + \frac{1}{r^2},\\
{T_r}^r &= - e^{-\lambda}\left(\frac{1}{r^2}
+ \frac{\nu^\prime}{r}\right) + \frac{1}{r^2},\\
{T_\theta}^\theta &= - \frac{e^{-\lambda}}{2}
\left(\nu^{\prime\prime}+
\frac{{\nu^\prime}^2}{2} + \frac{\nu^\prime-\lambda^\prime}{r} -
\frac{\nu^\prime\,\lambda^\prime}{2}\right), \\
{T_\phi}^\phi  &= {T_\theta}^\theta,
\end{align}
\end{subequations}
where the above $T^{\mu}_{\nu}$ components refer to the energy-momentum tensor, we can get several classes of solutions if  $\lambda = -\nu$. Among all, the standard vacuum solution by Schwarzchild, the Reisser-Nordstr\"{o}m solution for charged a BH and, last but not least, the de Sitter solution, corresponding to a uncharged singularity with a constant non-zero contribution to the energy-momentum tensor.

To be more general, assuming the above mathematical structure of our solution, we propose
\begin{equation}\label{lambda}
\lambda=-\nu=-\ln\left(1-\frac{2M}{r}-\Lambda r^2 + \frac{\mathcal C}{r^2}\right)\,,
\end{equation}
giving rise to a charged solution with vacuum energy term, \textit{i.e.}, with a non-zero constant value entering the energy momentum tensor. The spacetime therefore reads
\begin{align}
\nonumber
{\rm d}s^2 =\,& \displaystyle-\left(1-\frac{2 M}{r}-\Lambda r^2 + \frac{\mathcal C}{r^2}
\right)\,{\rm d}t^2+\\
\,&\left(1-\frac{2 M}{r}-\Lambda r^2 + \frac{\mathcal C}{r^2}
\right)^{-1} {\rm d}r^2 + r^2 {\rm d}\Omega^2,
\label{general solution}
\end{align}
where $M$ is the source mass, $\Lambda$ is the cosmological constant term that acts as energy of the ground state in the energy-momentum tensor, $\mathcal C$ is a constant related to the net charge $Q$, that reduces to Reissner-Nordstr\"{o}m to $\mathcal C\equiv Q^2$ and finally  ${\rm d}\Omega^2 \equiv {\rm d}\theta^2 + \sin^2\theta\, {\rm d}\phi^2$.

We consider \emph{three hierarchies} of the above metric:
\begin{itemize}
\item[-] \textit{model 1} with $\Lambda=\mathcal{C}=0$, leading to the simplest Schwarzschild solution;
\item[-] \textit{model 2} with  $\mathcal{C}=0$ but $\Lambda\neq0$, that turns out to be a de Sitter solution with the sign unspecified\footnote{In other words, we assume that both de Sitter (positive energy) and anti-de Sitter (negative energy) are possible.};
\item[-] \textit{model 3} with  $\Lambda\neq0$ and $\mathcal{C}\neq 0$, \textit{i.e.}, the full metric prompted in Eq. \eqref{general solution}.
 \end{itemize}

The physical meaning of the above scenarios is explained in the details below.
\begin{itemize}
\item[1)] We consider, in GR, the most general spherically symmetric spacetime, in a static configuration.
\item[2)] We test whether the energy momentum tensor provides or not a non-vanishing (constant) contribution associated with the compact object.
\item[3)] We explore the possibility that \emph{globally} NS sources exhibit a non-zero net charge, being not perfectly neutral from the outside, as a consequence of $\mathcal C\neq0$.
\end{itemize}

The main limitations of the above approach are:
\begin{itemize}
\item[a)] commonly speaking a NS is not perfectly static, but rather it is supposed to rotate;
\item[b)] the sign of $\Lambda$ is not fixed \emph{a priori}, leaving open the issue of being a de Sitter or an anti-de Sitter framework to describe the energy momentum tensor;
\item[c)] the NS, in general locally or globally neutral, may exhibit a non-specified net charge $Q$ showing slight deviation from the overall neutrality of the star.
\end{itemize}

Consequently, a more refined approach would include effects of rotation, providing an explanation about the sign of the rest energy,  specifying better how the charge is associated to the NS.

However, as a first toy-model treatment we motivate our choice since the standard Schwarzschild metric has been severely adopted to match the QPO data with encouraging results. So it appears natural to extend first the scenario of static configurations rather than working non-static frameworks out.

\subsection{QPOs from dynamics of test particles}

Bearing all the above in mind,  we start with the Lagrangian of the test particles
\begin{eqnarray}
{\mathbb{L}}=\frac{1}{2}m g_{\mu\nu} \dot{x}^{\mu} \dot{x}^{\nu} \ ,
\end{eqnarray}
in which $m$ is the test particle mass and $x_\mu(\tau)$  the worldlines. Here, $\dot{x}^{\mu}=dx^\mu/d\tau$ represents their four-velocities and so, since the metric is static the conserved quantities are the (specific) energy and angular momentum, namely $g_{tt}\dot{t}=-{\cal E}$ and $g_{\phi \phi}\dot{\phi} = {\cal L}$, being ${\cal E}$ and ${\cal L}$ the energy and the angular momentum per unit mass of the test particle, respectively.
Moreover, we introduce the normalization parameter $\epsilon=g_{\mu \nu}\dot{x}^{\mu} \dot{x}^{\nu}$ that describes null geodesics of massless particles for $\epsilon=0$, whereas corresponds to massive particles with time-like geodesics for $\epsilon=-1$.

As $m\neq0$, the equations of motion are
\begin{equation}
\label{eqmotionneutral}
\dot{t}=-\frac{{\cal E}}{g_{tt}}\ \ ,\ \  \dot{\phi}=\frac{{\cal L}}{g_{\phi \phi}}\ \ ,\ \ g_{rr} \dot{r}^2+g_{\theta\theta}\dot{\theta}^2=V_{\rm eff}\,,
\end{equation}
where the effective potential $V_{\rm eff}$ is
\begin{equation}
\label{effpotentail}
     V_{\rm eff}(r) = -\left(1+\frac{{\cal E}^2g_{\phi\phi}+{\cal L}^2g_{tt}}{g_{tt}g_{\phi\phi}}\right)
\end{equation}
For circular orbits in the equatorial plane one has $\dot{r}=\dot{\theta}=0$. Thus, the equations for orbital parameters are given as follows:
\begin{subequations}
\begin{eqnarray}
\Omega_{\phi}&=&\pm\sqrt{-\frac{{\partial_r}g_{tt}}{{\partial_r}g_{\phi\phi}}},\\
\dot{t}&=&u^t=\frac{1}{\sqrt{-g_{tt}-g_{\phi\phi}\Omega_\phi^2}},\\
{\cal E}&=&-\frac{g_{tt}}{\sqrt{-g_{tt}-g_{\phi\phi}\Omega_\phi^2}},\\
{\cal L}&=&\frac{g_{\phi\phi}\Omega_\phi}{\sqrt{-g_{tt}-g_{\phi\phi}\Omega_\phi^2}},
\end{eqnarray}
\end{subequations}
where the sign is $+$ ($-$) for co-rotating (counter-rotating) orbits \cite{2016EL....11630006B}.
The energy and angular momentum for the metric we have adopted are
\begin{subequations}
\begin{align}
   {\cal E}=&\,\frac{\mathcal C + r (r - r^3 \Lambda-2 m )}{r \sqrt{
 2 \mathcal C + r (r-3 m)}}\,\\
   {\cal L}=&\, \frac{r \sqrt{-\mathcal C + m r - r^4 \Lambda}}{\sqrt{2 \mathcal C + r (r-3 m)}}\,.
\end{align}
\end{subequations}

Thus, the fundamental frequencies of test particles around a compact object are easily computable and they can be converted to seconds, say ${\rm Hz}\equiv s^{-1}$, very easily in order to compare them with data.

In the regime of small oscillations, the displacements from equilibrium positions are  $r\sim r_0+\delta r$ and $\theta\sim\pi/2+\delta \theta$ and thus oscillations occur as
\begin{eqnarray}
\frac{d^2\delta r}{dt^2}+\Omega_r^2 \delta r=0\ , \qquad \frac{d^2\delta\theta}{dt^2}+\Omega_\theta^2 \delta\theta=0\ ,
\end{eqnarray}
with the frequencies
\begin{eqnarray}
\Omega_r^2&=&-\frac{1}{2g_{rr}(u^t)^2}\partial_r^2V_{\rm eff}(r,\theta)\Big |_{\theta=\pi/2}\,,\\
\Omega_\theta^2&=&-\frac{1}{2g_{\theta\theta}(u^t)^2}\partial_\theta^2V_{\rm eff}(r,\theta)\Big |_{\theta=\pi/2}\,,
\end{eqnarray}
for radial and angular oscillations, respectively.

So, adopting our spacetime, we get
\begin{subequations}
\begin{align}\label{wt}
\Omega_\phi^2=\,&\frac{m}{r^3} -\frac{\mathcal C}{r^4}- \Lambda,\\
\Omega_\theta^2 =\,&\Omega_\phi^2,\\
\nonumber
\Omega_r^2=\,&\frac{m(r-6 m)}{r^4} + \frac{\Lambda (15 m - 4 r)}{r} +\\
&\,\frac{3 \mathcal C (3 m - 4 r^3 \Lambda)}{r^5}-\frac{4\mathcal C^2}{r^6}\,,
\end{align}
\end{subequations}
where $\Omega_\phi$ represents the particle angular velocity measured by an asymptotic observer placed at infinity.

From the above angular frequencies, we define the Keplerian frequency $f_\phi=\Omega_\phi/(2\pi)$ and the radial epicyclic frequency of the Keplerian motion $f_r=\Omega_r/(2\pi)$.
The relativistic precession model identifies the lower QPO frequency $f_{\rm L}$ with the periastron precession, namely $f_{\rm L}=f_\phi-f_r$, and the upper QPO frequency $f_{\rm U}$ with the Keplerian frequency, namely $f_{\rm U}=f_\phi$.

Another physical quantity which is of great interest is the radius of the innermost circular stable orbit (ISCO), $r_{ISCO}$. It is determined from the following conditions
\begin{equation}
    \frac{d \cal E}{dr}=0\quad,\quad
    \frac{d \cal L}{dr}=0\,.
\end{equation}
For the Schwarzschild spacetime $r_{ISCO}=6M$, for the Reisser-Nordstr\"{o}m $r_{ISCO}$ is calculated in Ref.~\cite{2011PhRvD..83b4021P} and for other solutions $r_{ISCO}$ is cumbersome. Nevertheless, for our three hierarchies, we estimate $r_{ISCO}$ numerically instead of computing it analytically. This choice has been made due to the complexity of $r_{ISCO}$ with the involved  metrics. Thus, numerical results are reported in Tab. \ref{tab:isco}. As a matter of fact, these outcomes certify that model 3 appears mainly disfavored with respect to model 2, in general. Indeed, for two sources (Cir X-1 and 4U1614+091) ISCOs become un-physical, in agreement with large mass predictions got from MCMC analyses, clearly incompatible with current bounds on NS masses.

\section{The case of $F(R)$ theories}\label{sezione3}

The search for departures from GR represents an open challenge for both cosmology and astrophysics \cite{2010RvMP...82..451S}.

Alternatives to Einstein's gravity can be split into \emph{extended theories of gravity} and \emph{modified theories of gravity}, see e.g. \cite{2012PhR...513....1C}.
We here limit our attention to the extended class of models named $F(R)$, corresponding to analytical functions that extend the Ricci scalar in the gravitational Lagrangian, fulfilling the equivalence principle \cite{Sotiriou:2006hs}.

We motivate our choice, noting that the most recent Planck satellite results seem to indicate that the most suitable inflationary potential is conformally-invariant with $F(R)=R+\alpha R^2$, \textit{i.e.}, with the simplest polynomial extension to GR\footnote{Alternatively, a Higgs inflation, neglecting the kinetic term during inflation, is equivalent to $F(R)=R+\alpha R^2$ \cite{2021JKPS...78..897C}.} \cite{1980PhLB...91...99S}.

The gravitational action of $F(R)$-gravity is given by $
 I=\int d^4 x\sqrt{-g}\left[\frac{F(R)}{2}+\mathcal{L}_m\right]$, where $F(R)$ is a function of the Ricci scalar $R$ and $\mathcal{L}_m$ is the matter Lagrangian\footnote{Those theories may reproduce, although with several drawbacks, the cosmological epochs of inflation and dark energy \cite{2012Ap&SS.342..155B}. However, simpler versions making use of scalar fields in GR seem to be promising examples of unified dark energy models, see e.g. \cite{Capozziello:2018mds,Boshkayev:2019qcx,DAgostino:2022fcx,DAgostino:2021vvv,Luongo:2018lgy}, being consistent with low data \cite{Aviles:2016wel}.}.

The field equations become
\begin{align}
\nonumber
F_R(R)G_{\mu\nu}=&\,\kappa^2 T_{\mu\nu}+\frac{1}{2}g_{\mu\nu}\left[F(R)-RF_R(R)\right]
+\\
&\,(\nabla_{\mu}\nabla_{\nu}-g_{\mu\nu}\Box)\,F_R(R)\,,
\label{fieldequationF(R)}
\end{align}
where $F_{R}(R)\equiv dF(R)/dR$ and $G_{\mu\nu}\equiv R_{\mu\nu}-\frac{1}{2}g_{\mu\nu}R$, the Einstein tensor. For $F(R)\rightarrow R$, GR is recovered.

Starting from the considerations that in a spherically symmetric configuration in vacuum, differently from GR where one gets $R=0$ identically, the modified field equations produce the constraints $F(R_0)=0$, $F_R(R_0)=0$, with $R=R_0$  a real \emph{positive} or \emph{negative} constant.

This framework can account for different classes of models and, following Ref. \cite{2018EPJC...78..178C} and assuming a convenient $B=B(r)$, we write
\begin{eqnarray}\label{metric}
 ds^2=-Bdt^2+\frac{dr^2}{B}+\frac{r^2d\rho^2}{1-{k \rho^2}} +r^2\rho^2 d\phi^2.
\end{eqnarray}
So, from the above relation, considering the extended field equations, easily one can compute the Ricci scalar curvature that reads
\begin{align}
R=&\,-B'' - {4B'\over r}
- {2B\over r^{2}}+{2k\over r^2} \,,
\label{Ricci}
\end{align}
where $B^\prime=\partial_r B$ and $B^{\prime\prime}=\partial_{rr} B$. Thus, as $F(R)$-gravity leads to fourth order differential equations of motion, assuming $R=R_{0}$, we recover a  second order differential equation whose solution depends upon  two integration constants. So, taking the simplest case  $R=R_{0}$ we get
\begin{equation}
B(r)=k-\frac{c_0}{r}+\frac{c_1}{r^2}- \frac{R_0}{12}r^2\,,
  \label{ex1}
\end{equation}

\subsection{Degeneracy problem}

In the above picture, the free parameters are $k, c_0, c_1$ and $R_0$. It appears clear that
$c_1>0$ provides a (topological)
Reissner-Nordstr\"{o}m--de Sitter or anti-de Sitter, respectively for $R_0>0$ and $R_0<0$. In this picture, however, $c_1$ does not correspond exactly to the charge of
an \emph{external electric field}, but it appears as a consequence of the complexity of the fourth order theories of gravity that we are considering. Say it differently, $c_1$ turns out to be an integration constant for the vacuum solution from which we started in $F(R)$ theories.

This spacetime, for $k=1$, fully recovers the broadly general class of metrics that we have proposed in Eq. \eqref{general solution}.
There, the net charge is exactly provided by an external electic field, while $\Lambda$ is here reexpressed by virtue of the constant Ricci scalar, $R_0$. In other words, the two metrics strongly degenerate, albeit the physics associated with them may differ.
On the one hand,  at high distances, the case $\Lambda\rightarrow0$ immediately provides that $k=1$ without the need of further cases, $k\neq1$. So, without losing generality, it appears evident that it can be fixed as $k=1$, having that the metric can act as a BH, once the event horizon is located for a given  real positive value of radius, $r_{H}$, with $B(r)$ to vanish.
On the other hand, the free $c_0$ term is positive definite in order to avoid repulsive gravity effects \cite{un,due,tre,quattro}, so that one can conclude
\begin{subequations}
\begin{align}
&k=1\,,\\
&c_0=2M\,,\\
&c_1=Q^2\,,\\
&R_0=12\Lambda\,,
\end{align}
\end{subequations}
in order to obtain back our previous solution.

However, for stable horizons, $c_1<0$ seems to be required\footnote{Essentially this comes from the requirement that positive roots of $B(r)$ are necessary}, while, in principle, one can adopt a different topology in $F(R)$, say
$k=0\,,-1$.

An interesting point is also that $R_0\neq 0$ has no specified sign, leading to a de Sitter or anti-de Sitter solution\footnote{In extended theories of gravity, this appears easier to account as one needs anti-de Sitter with  $R_0 < 0$, choosing $k = -1$. This would  preserve the metric signature. In GR, a different sign of $\Lambda$ implies different properties of the energy ground state.}. However, in GR this appears tricky: choosing anti-de Sitter instead of de Sitter would modify the physics associated with the NS net energy.

Consequently, our outcomes would indicate whether robust departures from the physical expectations of GR can be found adopting our eight sources and Eq. \eqref{general solution}.

\section{Monte Carlo analysis and theoretical discussion}\label{sezione4}

\begin{table*}
\centering
\setlength{\tabcolsep}{.5em}
\renewcommand{\arraystretch}{1.3}
\begin{tabular}{lcccrrrrr}
\hline\hline
Source                                  &
$M$                                     &
$R_0$                                   &
$\mathcal C$                            &
$\ln L_{\rm max}$               &
AIC                                     &
BIC                                     &
$\Delta$AIC                             &
$\Delta$BIC                             \\
                                        &
$({\rm M}_\odot)$                       &
$(\times10^{-5}\,{\rm km}^{-2})$        &
$({\rm km}^2)$                          & & & & &\\
\hline
                                        &
$2.224^{+0.029\,(+0.058)}_{-0.029\,(-0.058)}$ &
--                                      &
--                                      &
$-125.84$                               &
$254$ & $254$ & $117$ & $115$ \\
Cir X1                                   &
$1.846^{+0.045\,(+0.091)}_{-0.045\,(-0.090)}$ &
$1.28^{+0.12\,(+0.23)}_{-0.12\,(-0.24)}$  &
--                                      &
$-70.07$                                &
$144$ & $145$ & $7$ & $6$\\
                                        &
$>3.00$ &
$1.17^{+0.12\,(+0.25)}_{-0.12\,(-0.24)}$  &
$37.67^{+0.81\,(+\ \,1.64)}_{-3.50\,(-14.01)}$   &
$-65.65$                                &
$137$ & $139$ & $0$ & $0$ \\
\hline
                                        &
$2.161^{+0.010\,(+0.020)}_{-0.010\,(-0.021)}$ &
--                                      &
--                                      &
$-200.33$ &
$403$ & $404$ & $187$ & $186$ \\
GX 5--1                                   &
$2.397^{+0.019\,(+0.038)}_{-0.019\,(-0.038)}$ &
$-6.46^{+0.48\,(+0.95)}_{-0.48\,(-0.95)}$  &
--                                      &
$-106.08$
& $216$ & $218$ & $0$ & $0$\\
                                        &
$1.96^{+0.78\,(+1.09)}_{-0.68\,(-0.91)}$ &
$-6.03^{+0.50\,(+0.93)}_{-1.09\,(-1.96)}$  &
$-7.10^{+13.7\,(+20.5)}_{-7.50\,(-9.0)}$  &
$-105.94$ &
$218$ & $221$ & $2$ & $3$\\
\hline
                                        &
$2.07678^{+0.0002\,(+0.0003)}_{-0.0003\,(-0.0007)}$  &
-- &
-- &
$-1819.02$ &
$3640$ & $3641$ & $3543$ & $3543$ \\
GX 17+2                                  &
$1.733^{+0.011\,(+0.021)}_{-0.011\,(-0.022)}$  &
$21.53^{+0.45\,(+0.91)}_{-0.45\,(-0.90)}$ &
-- &
$-46.42$ &
$97$ & $98$ & $0$ & $0$ \\
                                        &
$1.61^{+1.28\,(+1.32)}_{-0.57\,(-0.58)}$  &
$21.13^{+1.52\,(+2.03)}_{-3.30\,(-3.82)}$ &
$-1.54^{+20.34\,(+21.22)}_{-\ \,5.18\,(-\ 5.30)}$ &
$-46.42$ &
$99$ & $100$ & $2$ & $2$\\
\hline
                                        &
$2.102^{+0.003\,(+0.007)}_{-0.003\,(-0.007)}$  &
-- &
-- &
$-130.86$ &
$264$ & $264$ & $8$ & $7$ \\
GX 340+0                                 &
$2.149^{+0.015\,(+0.030)}_{-0.015\,(-0.031)}$  &
$-1.39^{+0.45\,(+0.89)}_{-0.44\,(-0.89)}$ &
-- &
$-126.06$ &
$256$ & $257$ & $0$ & $0$ \\
                                        &
$1.84^{+0.62\,(+1.28)}_{-0.59\,(-0.78)}$  &
$-1.24^{+0.42\,(+0.84)}_{-0.63\,(-2.08)}$ &
$-4.62^{+10.07\,(+23.59)}_{-\ \,6.19\,(-\ 6.99)}$ &
$-125.95$ &
$258$ & $259$ & $2$ & $2$\\
\hline
                                        &
$1.9649^{+0.0005\,(+0.0011)}_{-0.0005\,(-0.0011)}$ &
--    &
--      &
$-3887.17$   &
$7776$ & $7778$ & $7457$ & $7453$\\
Sco X1                                  &
$1.690^{+0.003\,(+0.007)}_{-0.003\,(-0.007)}$ &
$21.77^{+0.24\,(+0.49)}_{-0.25\,(-0.49)}$    &
--      &
$-158.61$   &
$321$ & $326$ & $2$ & $1$ \\
                                        &
$2.229^{+0.005\,(+0.010)}_{-0.038\,(-0.161)}$ &
$21.88^{+0.24\,(+0.54)}_{-0.22\,(-0.48)}$    &
$7.60^{+0.06\,(+0.09)}_{-0.59\,(-2.48)}$      &
$-156.42$   &
$319$ & $325$ & $0$ & $0$  \\
\hline
                                        &
$1.960^{+0.004\,(+0.007)}_{-0.004\,(-0.008)}$ &
-- &
--    &
$-235.83$   &
$474$ & $474$ & $342$ & $340$\\
4U1608--52                                   &
$1.728^{+0.014\,(+0.028)}_{-0.014\,(-0.028)}$ &
$17.62^{+0.94\,(+1.87)}_{-0.94\,(-1.88)}$ &
--    &
$-66.14$   &
$136$ & $137$ & $4$ & $3$ \\
                                        &
$3.057^{+0.018\,(+0.033)}_{-0.152\,(-2.267)}$ &
$13.43^{+1.51\,(+4.25)}_{-0.48\,(-1.90)}$ &
$22.19^{+0.46\,(+\ \,0.76)}_{-3.03\,(-30.11)}$    &
$-63.22$   &
$132$ & $134$ & $0$ & $0$\\
\hline
                                        &
$1.734^{+0.003\,(+0.006)}_{-0.003\,(-0.006)}$ &
-- &
--   &
$-212.61$   &
$427$ & $427$ & $353$ & $353$ \\
4U1728--34                                   &
$1.445^{+0.016\,(+0.032)}_{-0.016\,(-0.032)}$ &
$30.74^{+1.58\,(+3.15)}_{-1.58\,(-3.18)}$ &
--    &
$-35.15$   &
$74$ & $74$ & $0$ & $0$ \\
                                        &
$2.54^{+0.10\,(+0.11)}_{-1.55\,(-1.56)}$ &
$28.23^{+4.79\,(+6.68)}_{-2.02\,(-3.66)}$ &
$15.41^{+\ \,1.86\,(+\ \,1.96)}_{-19.45\,(-19.56)}$    &
$-34.95$   &
$76$ & $76$ & $2$ & $2$ \\
\hline
                                        &
$1.904^{+0.001\,(+0.003)}_{-0.001\,(-0.003)}$ &
-- &
--    &
$-842.97$   &
$1688$ & $1689$ & $1354$ & $1351$ \\
4U0614+091                                &
$1.545^{+0.011\,(+0.021)}_{-0.011\,(-0.021)}$ &
$28.39^{+0.80\,(+1.59)}_{-0.80\,(-1.59)}$ &
--    &
$-188.70$   &
$381$ & $384$ & $47$ & $46$ \\
                                        &
$>3.50$ &
$19.67^{+0.55\,(+1.71)}_{-0.89\,(-2.34)}$ &
$33.291^{+0.022\,(+0.174)}_{-0.537\,(-2.063)}$    &
$-163.99$   &
$334$ & $338$ & $0$ & $0$  \\
\hline
\end{tabular}
\caption{Best-fit parameters with the associated $1$--$\sigma$ ($2$--$\sigma$) error bars. For each source, the first, the second and the third lines list the results of the MCMC fits for model 1, 2 and 3, respectively. $\Delta$AIC and $\Delta$BIC are computed with respect to the reference model, \textit{i.e.}, the model with the highest value of $\ln \mathcal L_{\rm max}$.}
\label{tab:results}
\end{table*}

\begin{table}
\centering
\setlength{\tabcolsep}{1.em}
\renewcommand{\arraystretch}{1.1}
\begin{tabular}{lcccc}
\hline
\hline
Source & Model & ISCO &  Inner &  Outer \\
       &       & (km) &  (km)  &  (km) \\
\hline
Cir X1 & 1 & $19.62$ & $30.79$    &  $52.16 $  \\
$\,$ &  $2$ & $16.32$  & $28.84$ & $48.29$\\
$\,$ & $3$ & $-$  & $30.39$ & $53.65$\\
\hline
GX 5-1 & $1$ & $19.06$    &  $21.33$ & $31.70$   \\
$\,$ &  $2$ & $20.73$  & $22.15$ & $33.35$\\
$\,$ & $3$ & $20.16$  & $21.47$ & $31.93$\\
\hline
GX 17+2 & $1$ & $18.32$    &  $18.33$ & $22.94$   \\
$\,$ &  $2$ & $16.01$  & $17.04$ & $21.09$\\
$\,$ & $3$ & $15.89$  & $16.85$ & $20.81$\\
\hline
GX 340+0 & $1$ & $18.54$    &  $21.52$ & $29.07$   \\
$\,$ &  $2$ & $18.88$  & $21.71$ & $29.34$\\
$\,$ & $3$ & $18.50$  & $21.15$ & $28.44$\\
\hline
Sco X1 & $1$ & $17.33$    &  $17.72$ & $20.98$   \\
$\,$ &  $2$ & $15.60$  & $16.66$ & $19.58$\\
$\,$ & $3$ & $16.23$  & $17.43$ & $20.64$\\
\hline
4U1608-52 & $1$ & $17.29$    &  $17.65$ & $21.75$   \\
$\,$ &  $2$ & $15.82$  & $16.77$ & $20.50$\\
$\,$ & $3$ & $16.83$  & $18.30$ & $22.97$\\
\hline
4U1728-34 & $1$ & $15.30$    &  $16.06$ & $18.93$   \\
$\,$ &  $2$ & $13.37$  & $14.91$ & $17.43$\\
$\,$ & $3$ & $14.06$  & $16.35$ & $19.46$\\
\hline
4U0614+091 & $1$ & $16.80$    &  $16.95$ & $20.05$   \\
$\,$ &  $2$ & $14.34$  & $15.60$ & $18.30$\\
$\,$ & $3$ & $-$  & $17.66$ & $21.45$\\
\hline
\end{tabular}
\caption{Numerical values of ISCO and inner and outer disk radii for each source, computed from best-fit results of Tab.~\ref{tab:results}. Model 3 appears to be disfavored with respect to model 2, that turns out to be optimal. For two sources, namely Cir X1 and 4U1614+091, the ISCO of model 3 does not provide physical results. }
\label{tab:isco}
\end{table}

We perform a MCMC analysis by means of the Metropolis-Hastings algorithm, searching for the best-fit parameters that maximize the log-likelihood.
\begin{equation}
\label{loglike}
    \ln L = -\sum_{k=1}^{N}\left\{\dfrac{\left[f_{\rm L}^k-f_{\rm L}(p,f_{\rm U}^k)\right]^2}{2(\sigma f_{\rm L}^k)^2} + \ln(\sqrt{2\pi}\sigma f_{\rm L}^k)\right\}
\end{equation}
with $N$ data for each source, sampled as lower frequencies $f_{\rm L}^k$, attached errors $\sigma f_{\rm L}^k$, and upper frequencies $f_{\rm U}^k$.
The theoretically-computed frequencies $f_{\rm L}(p,f_{\rm U}^k)$ depend also on combinations of the parameters $p=\{M,\,R_0,\,\mathcal C\}$, depending on the considered scenario.

We modify the \texttt{Wolfram Mathematica} code from Ref.~\cite{2019PhRvD..99d3516A} and adapt it to the case of QPO data computing the widest possible parameter spaces over the free coefficients of our metric\footnote{Several applications of this code in cosmology, with different log-likelihoods, have been performed, e.g., in Refs.~\cite{Luongo:2020hyk,Luongo:2020aqw,Muccino:2020gqt,Luongo:2021pjs}.}. In particular, we consider the following priors over the coefficients\begin{subequations}
\begin{align}
M&\in [0 ; 5]\,{\rm M}_\odot\,,\\
R_0&\in [-50 ; 50]\times 10^{-5}\,{\rm km}^{-2}\,,\\
\mathcal C&\in [-60 ; 60]\,{\rm km}^2\,.
\end{align}
\end{subequations}

For each source we perform:
\begin{itemize}
    \item[-] three MCMC analyses for each hierarchy;
    \item[-] the computation of the log-likelihood from $\mathcal N\simeq 10^5$ total number of  iterations;
    \item[-] the search of the best-fit parameters providing the absolute and real maximum of the log-likelihood;
    \item[-] contours and statistical analyses of the errors, displayed up to 2--$\sigma$ confidence levels.
\end{itemize}

To assess the best-fit model out of the three scenarios derived from the underlying metric, we use the Aikake and the Bayesian Information Criterion, respectively AIC and BIC \cite{2007MNRAS.377L..74L}. Thus, considering our likelihood function from Eq.~\eqref{loglike}, we define
\begin{subequations}
\begin{align}
{\rm AIC}&=-2\ln L_{\rm max}+2p\,,\\
{\rm BIC}&=-2\ln L_{\rm max}+p\ln N\,,
\end{align}
\end{subequations}
where $\ln L_{\rm max}$ is the maximum value of the log-likelihood, $p$ the number of estimated parameters in the model and $N$ the number of the sample size. Recognizing the model with the lowest value of the AIC and BIC tests, say AIC$_{f}$ and BIC$_{f}$, as the fiducial (best-suited) model, the statistical evidence in support of the reference model is underlined by the difference $\Delta{\rm AIC/}\Delta{\rm BIC}={\rm AIC/BIC}-{\rm AIC/BIC}_f$.
Precisely, when comparing models, the evidence against the proposed model or, equivalently, in favor of the reference model can be naively summarized as follows:
\begin{itemize}
    \item $\Delta{\rm AIC}$ and $\Delta{\rm BIC}\in[0,\,3]$, weak evidence;
    \item $\Delta{\rm AIC}$ and $\Delta{\rm BIC}\in (3,\,6]$, mild evidence;
    \item $\Delta{\rm AIC}$ and $\Delta{\rm BIC}>6$, strong evidence.
\end{itemize}

\begin{figure}
\centering
\includegraphics[width=0.49\hsize,clip]{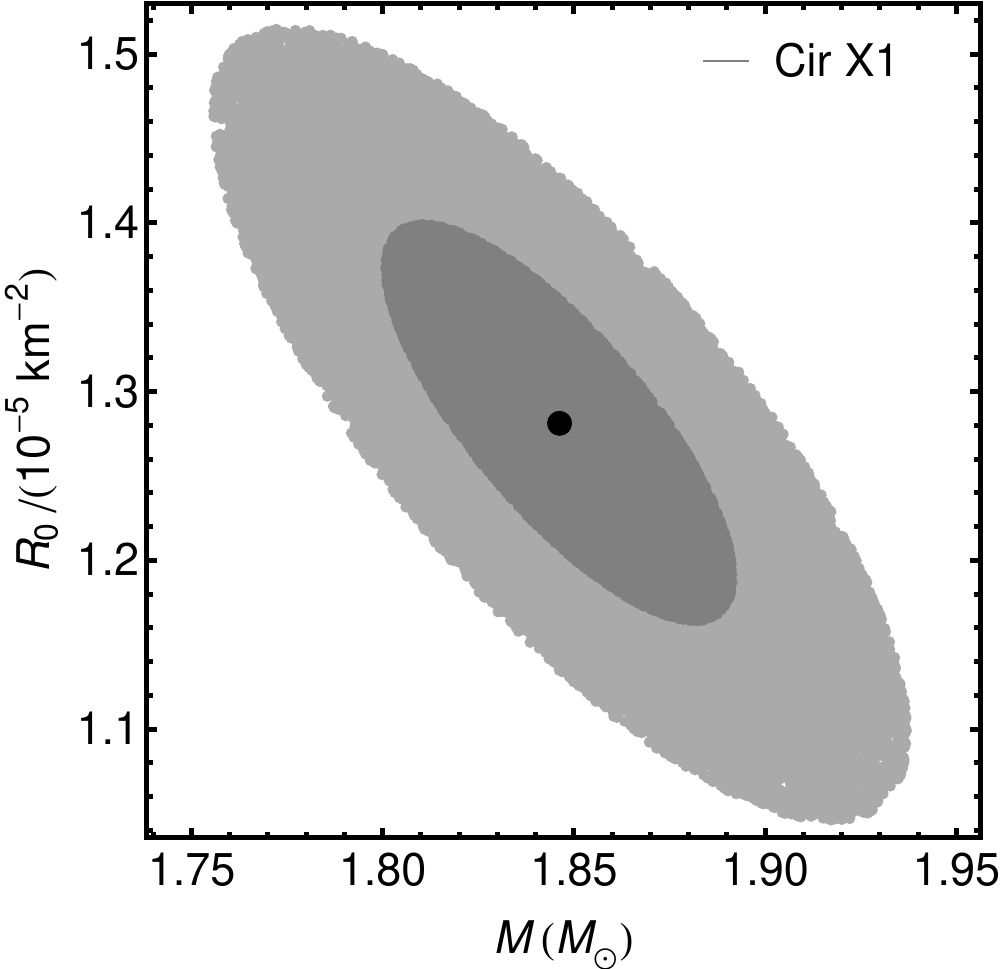}
\includegraphics[width=0.49\hsize,clip]{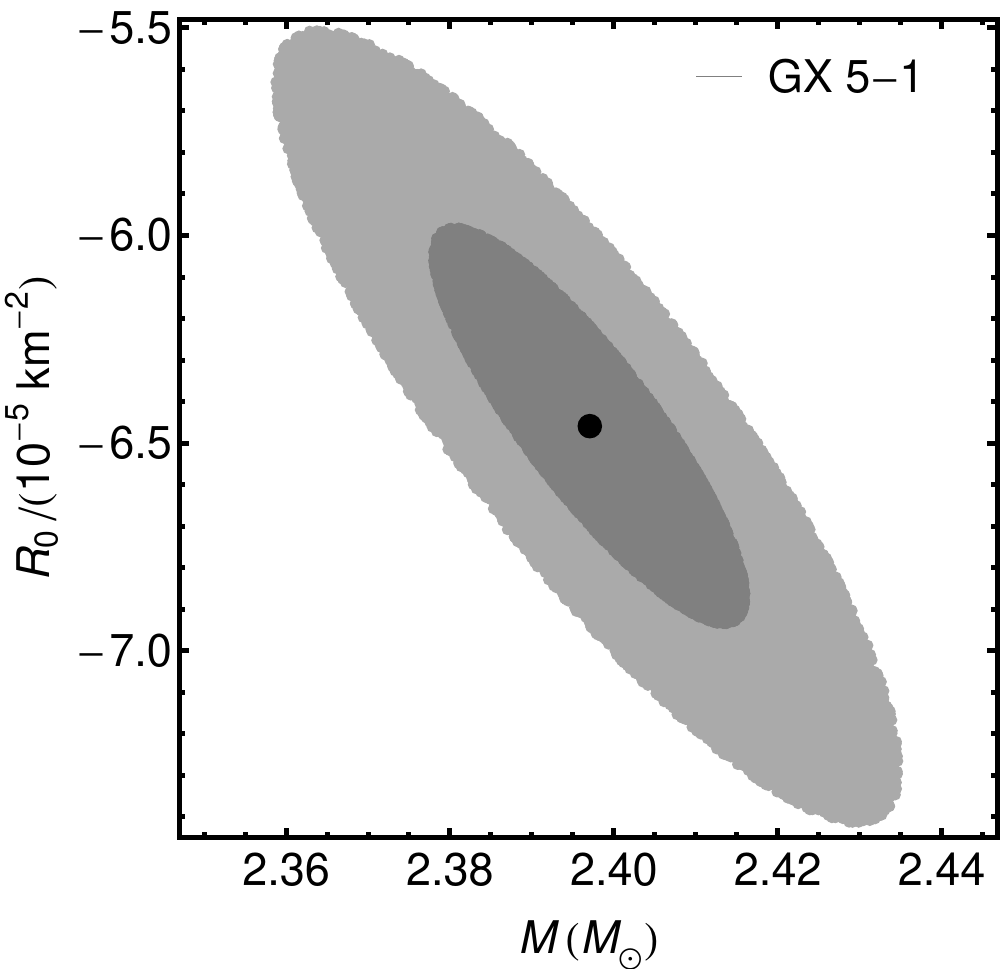}
\includegraphics[width=0.49\hsize,clip]{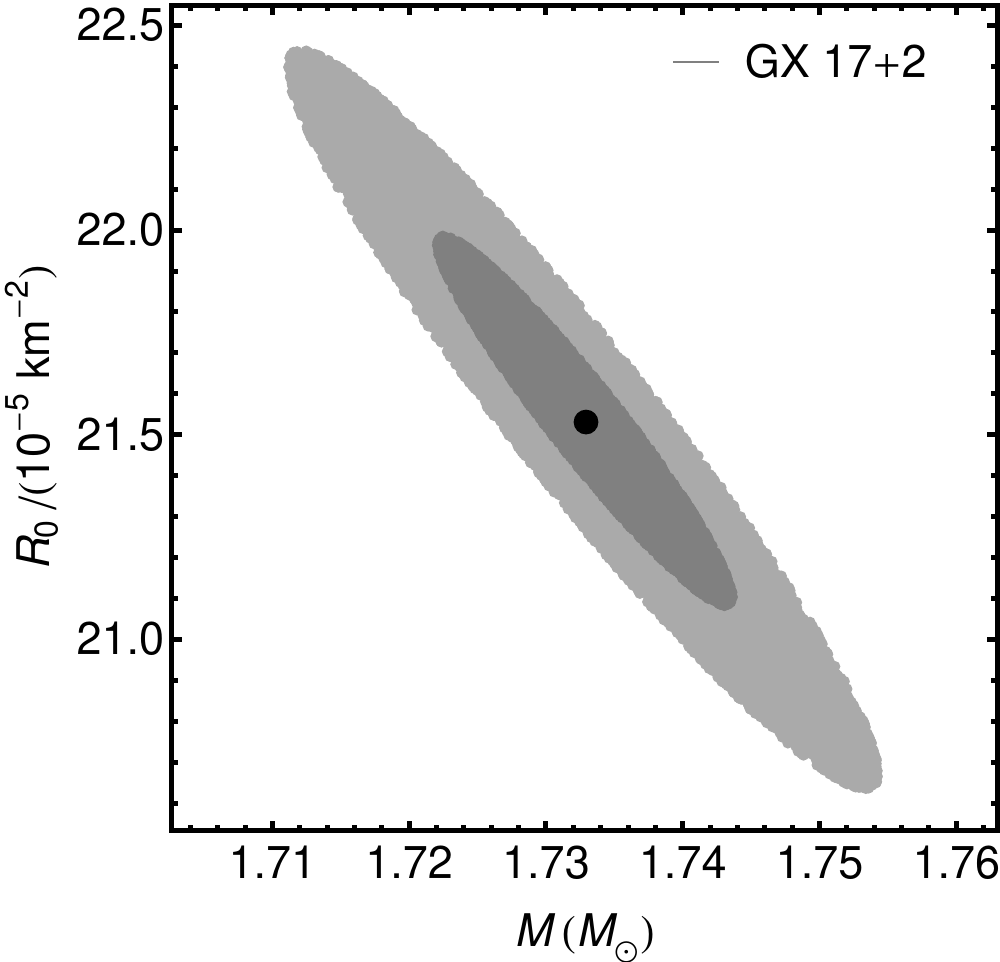}
\includegraphics[width=0.49\hsize,clip]{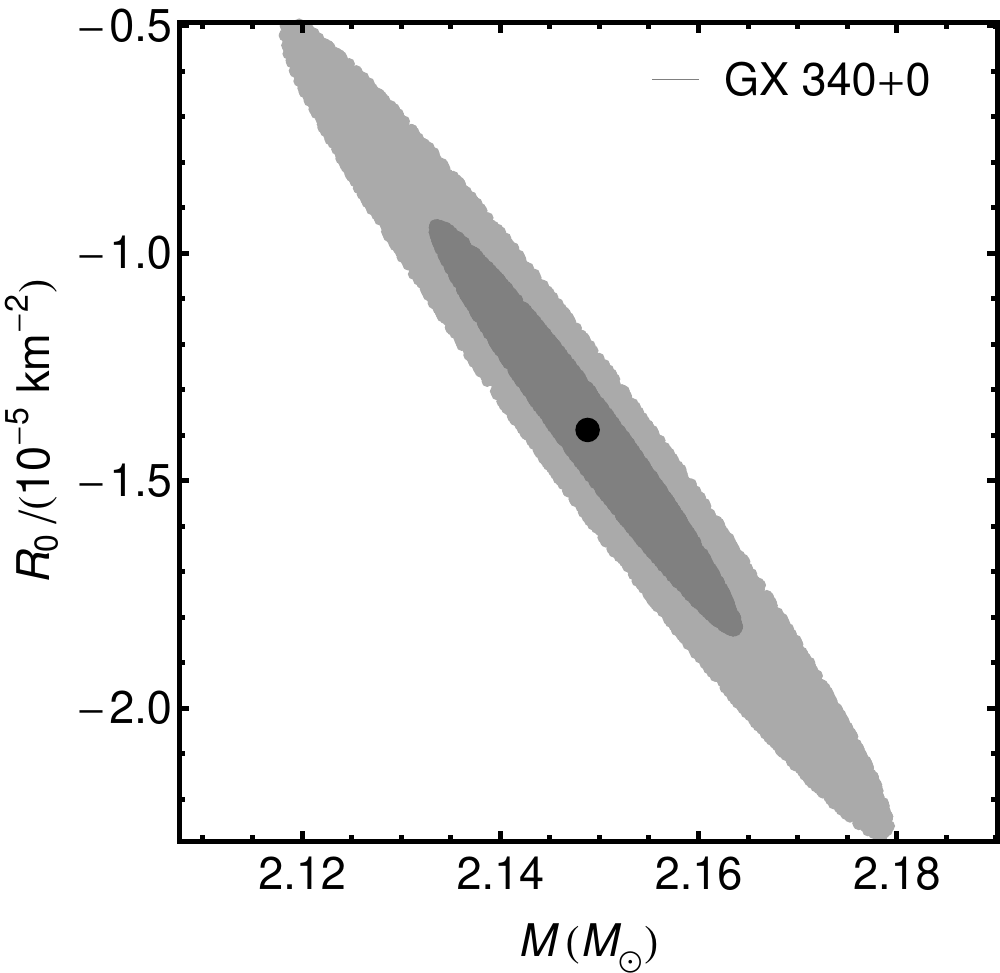}
\includegraphics[width=0.50\hsize,clip]{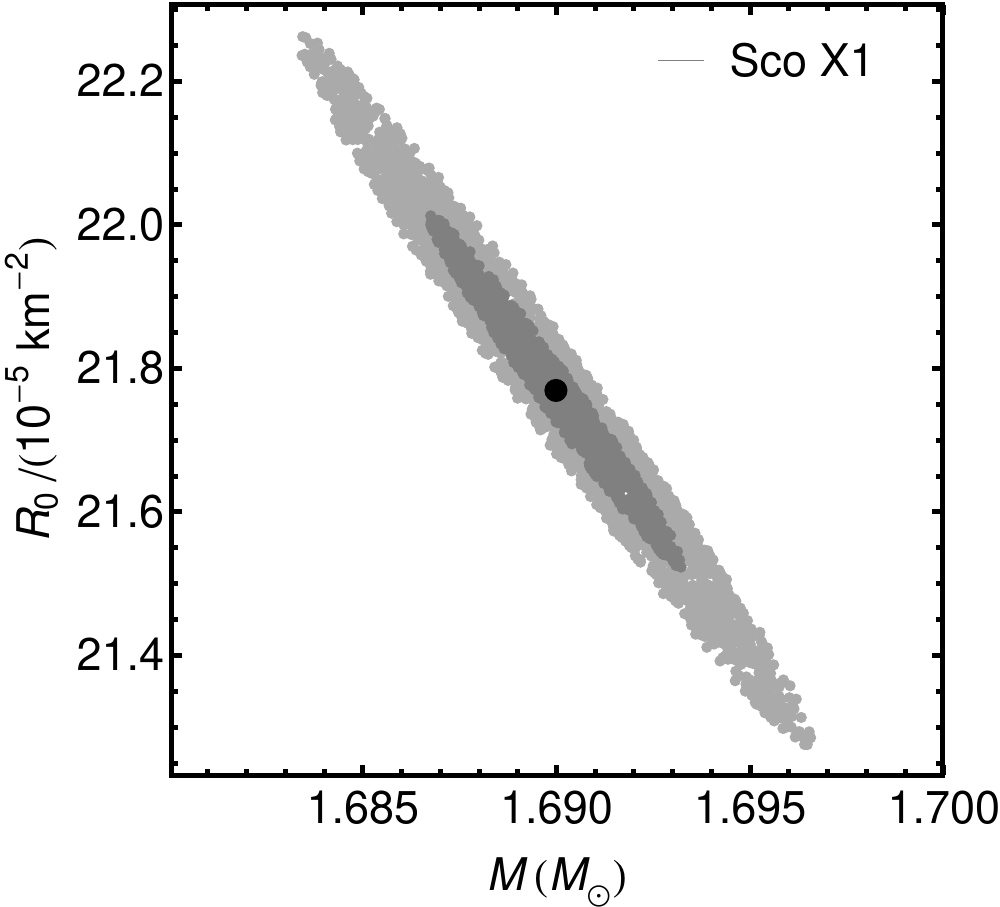}
\includegraphics[width=0.48\hsize,clip]{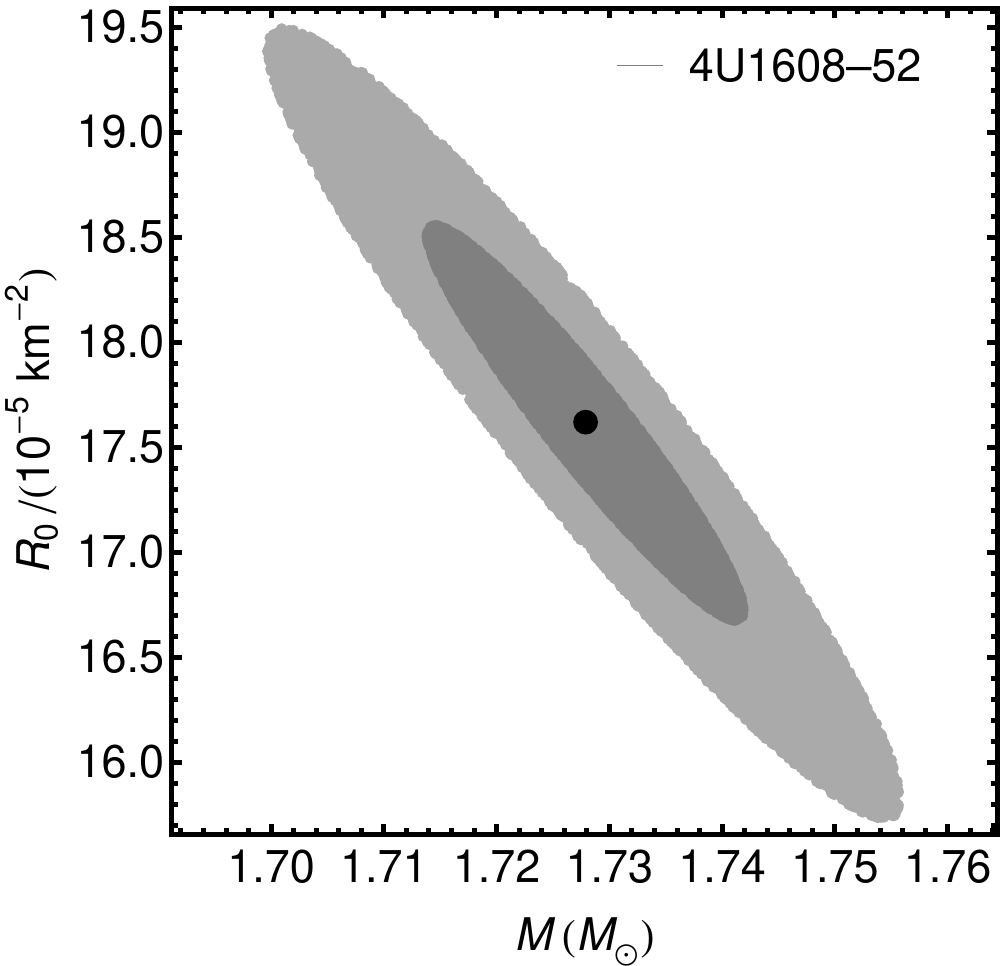}
\includegraphics[width=0.48\hsize,clip]{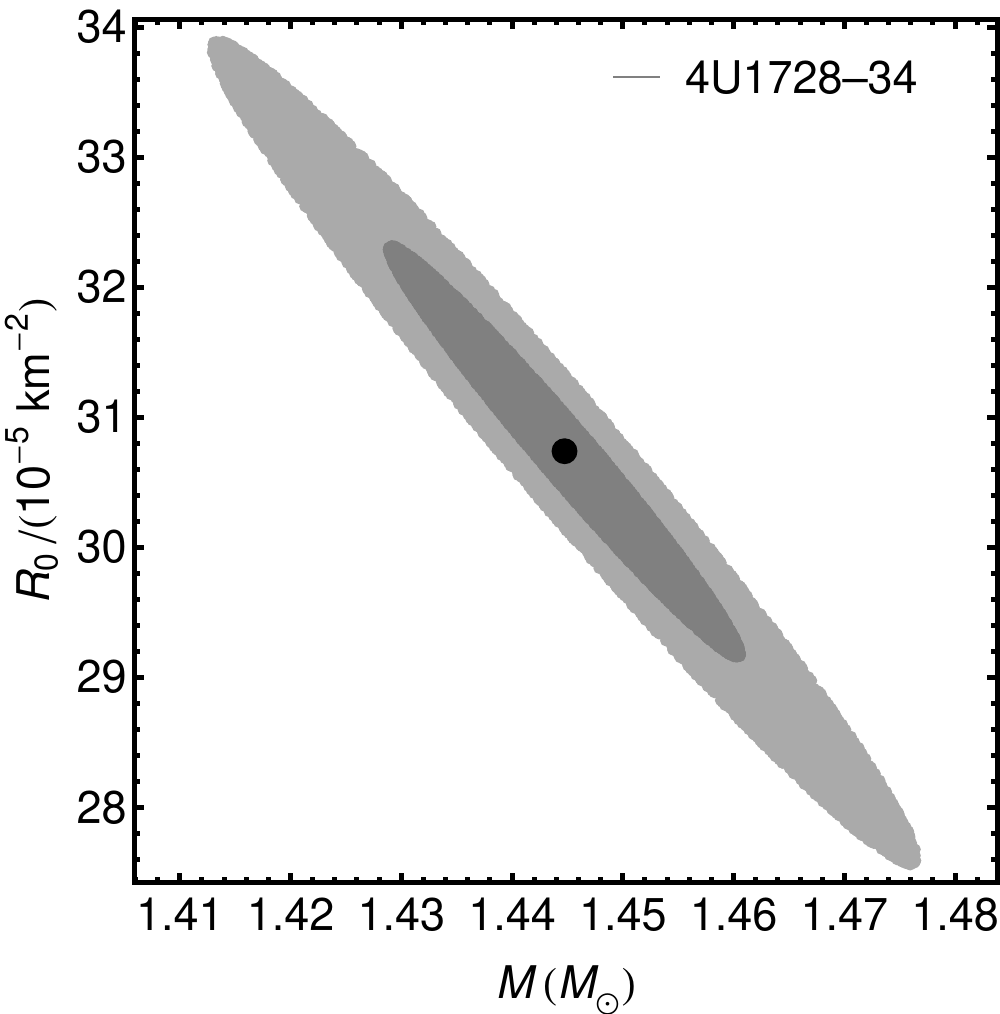}
\includegraphics[width=0.50\hsize,clip]{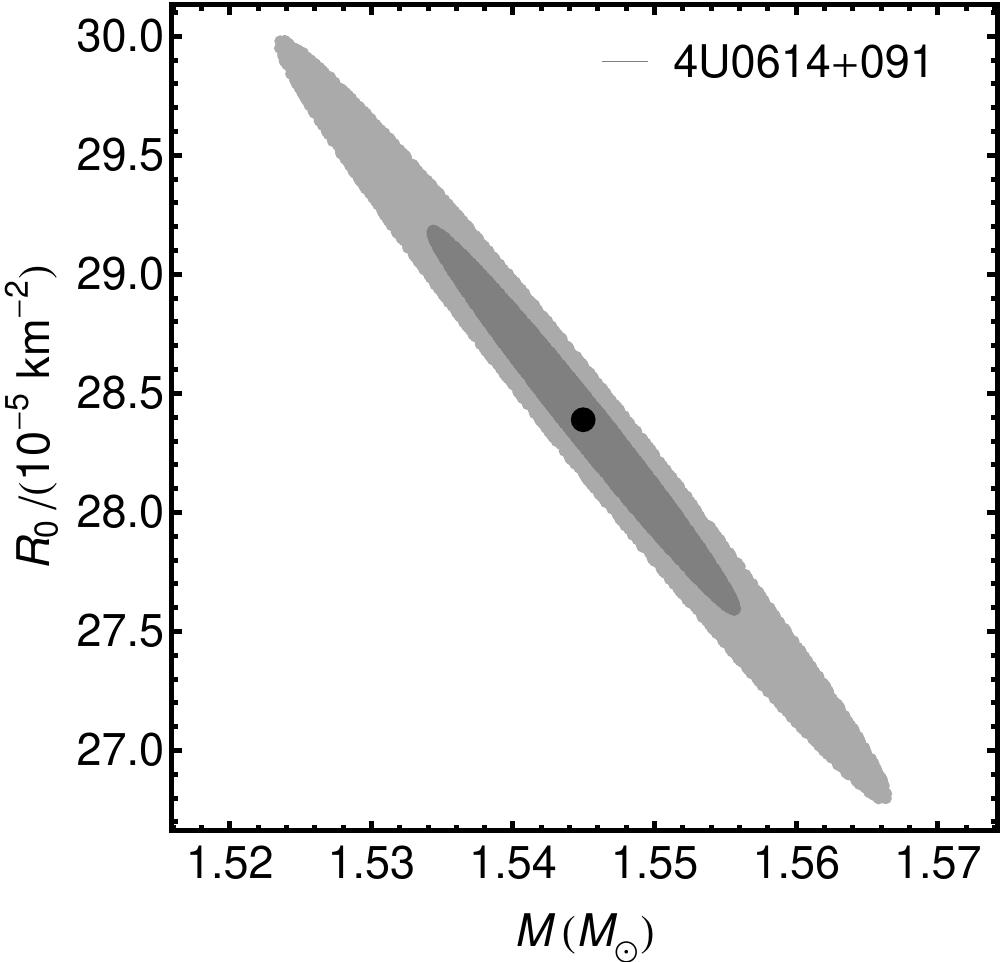}
\caption{Contours plots of the best-fit parameters (black dots) and the associated 1--$\sigma$ (dark gray) and 2--$\sigma$ (light gray) confidence regions of the sources listed in Tab. \ref{tab:results}.}
\label{fig:contours}
\end{figure}

\begin{figure*}
{\hfill
\includegraphics[width=0.42\hsize,clip]{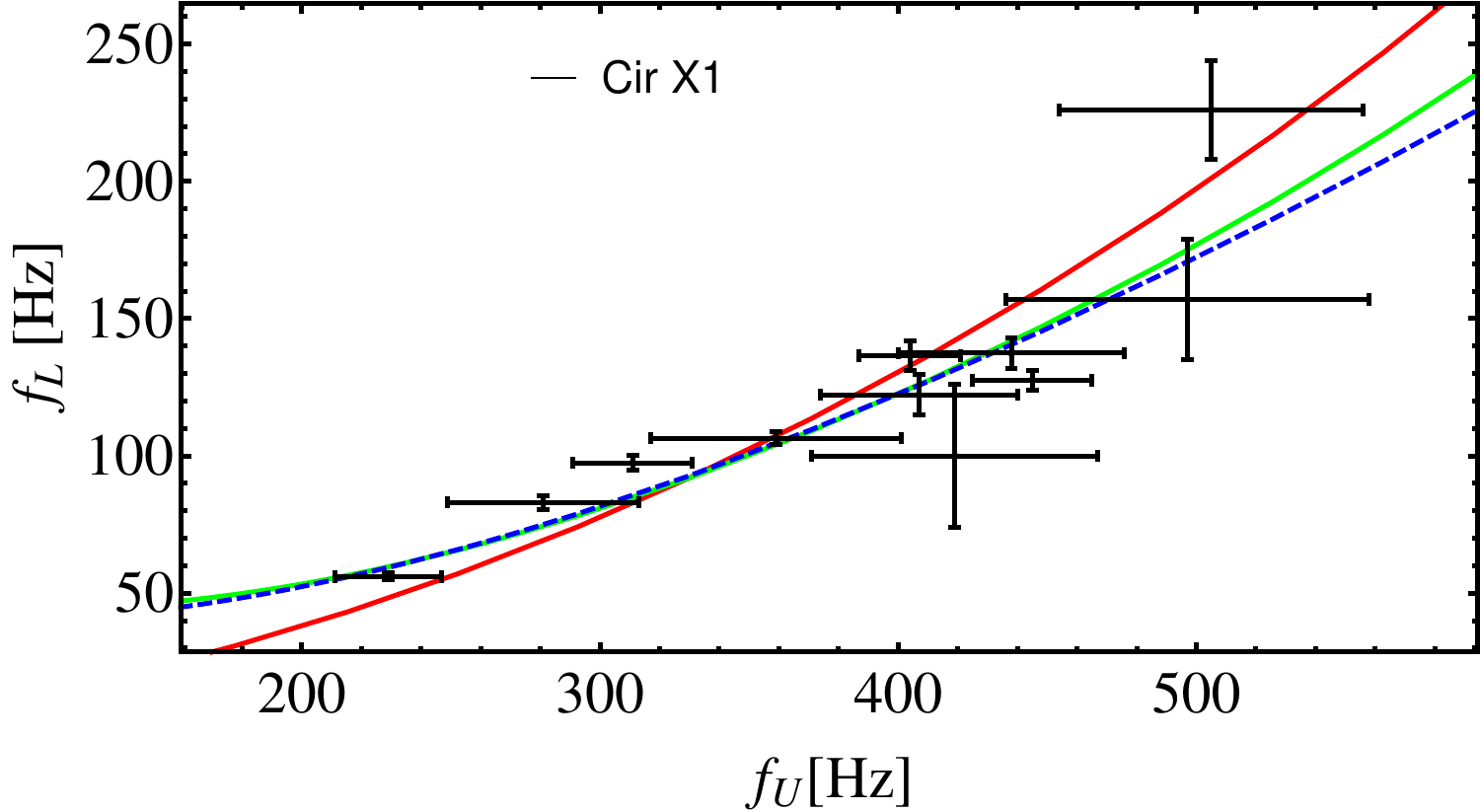}\hfill
\includegraphics[width=0.42\hsize,clip]{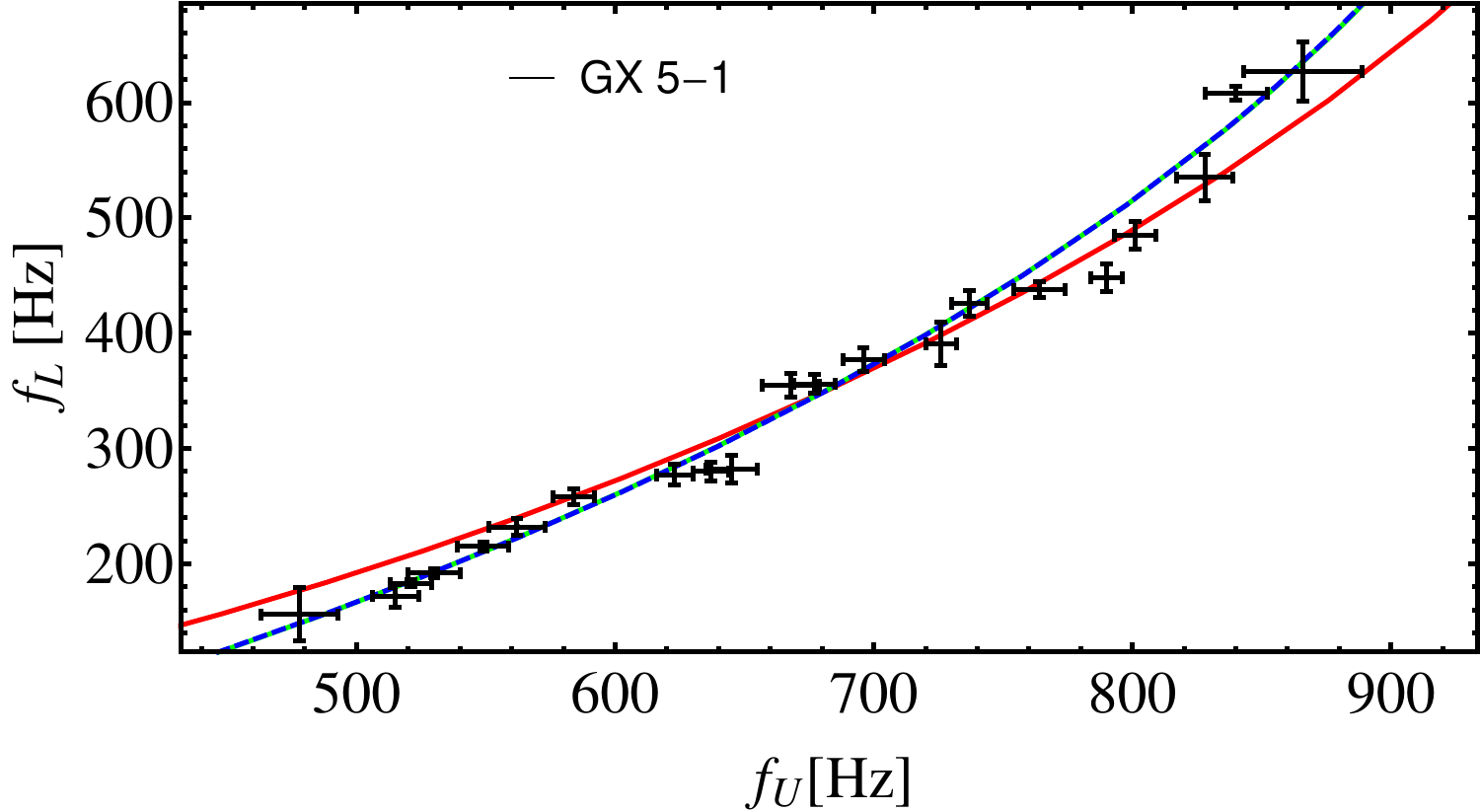}\hfill}

{\hfill
\includegraphics[width=0.42\hsize,clip]{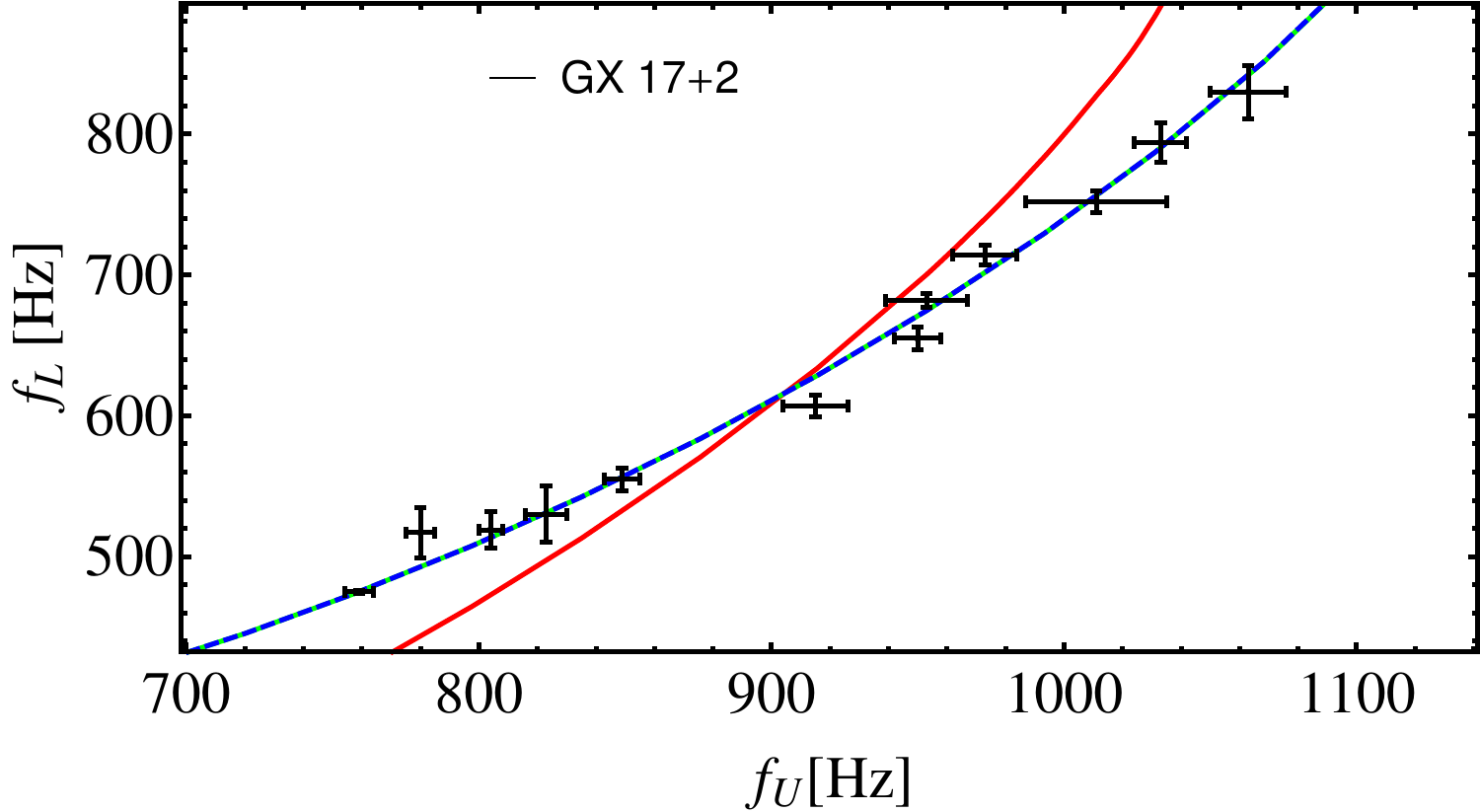}\hfill
\includegraphics[width=0.42\hsize,clip]{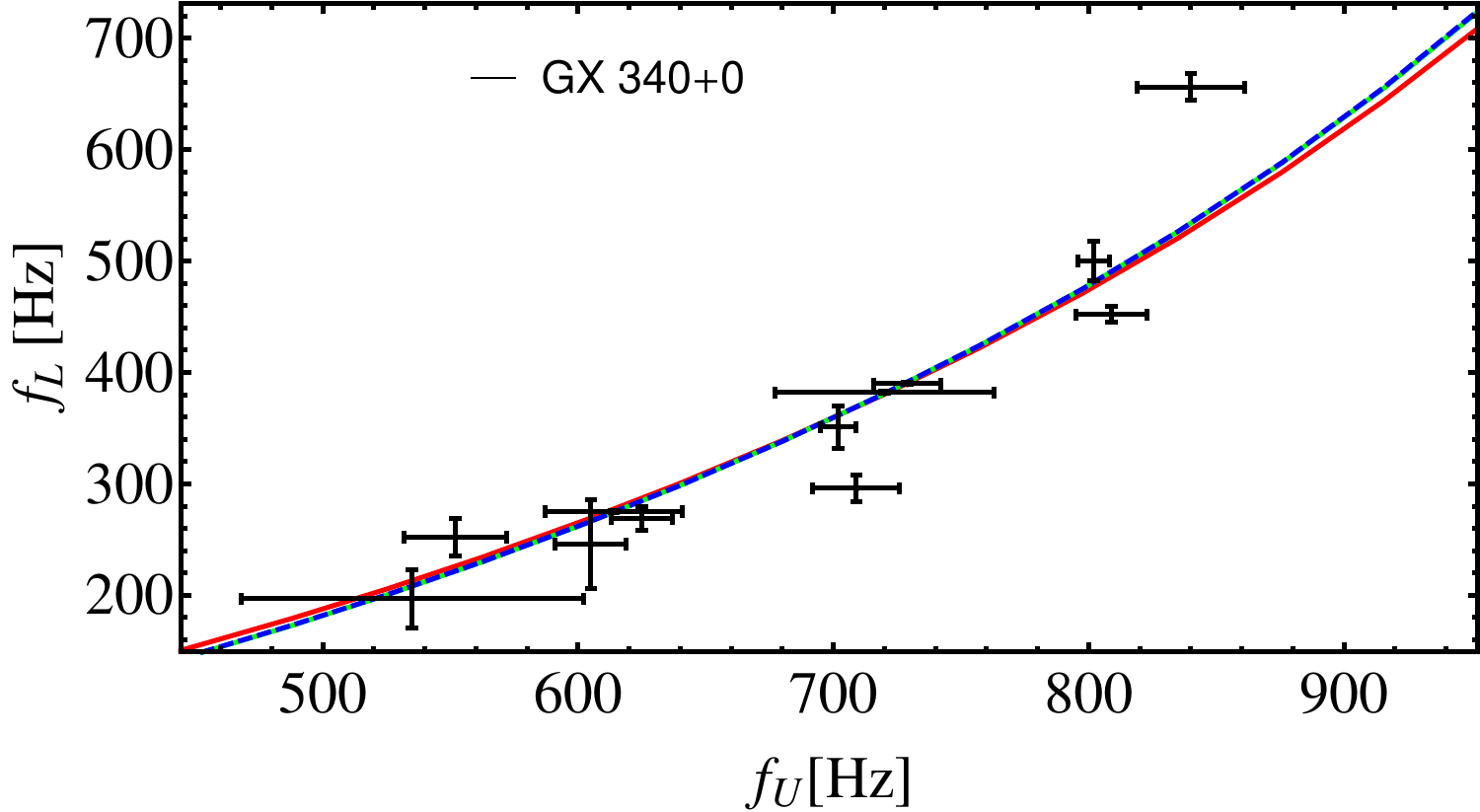}\hfill}

{\hfill
\includegraphics[width=0.42\hsize,clip]{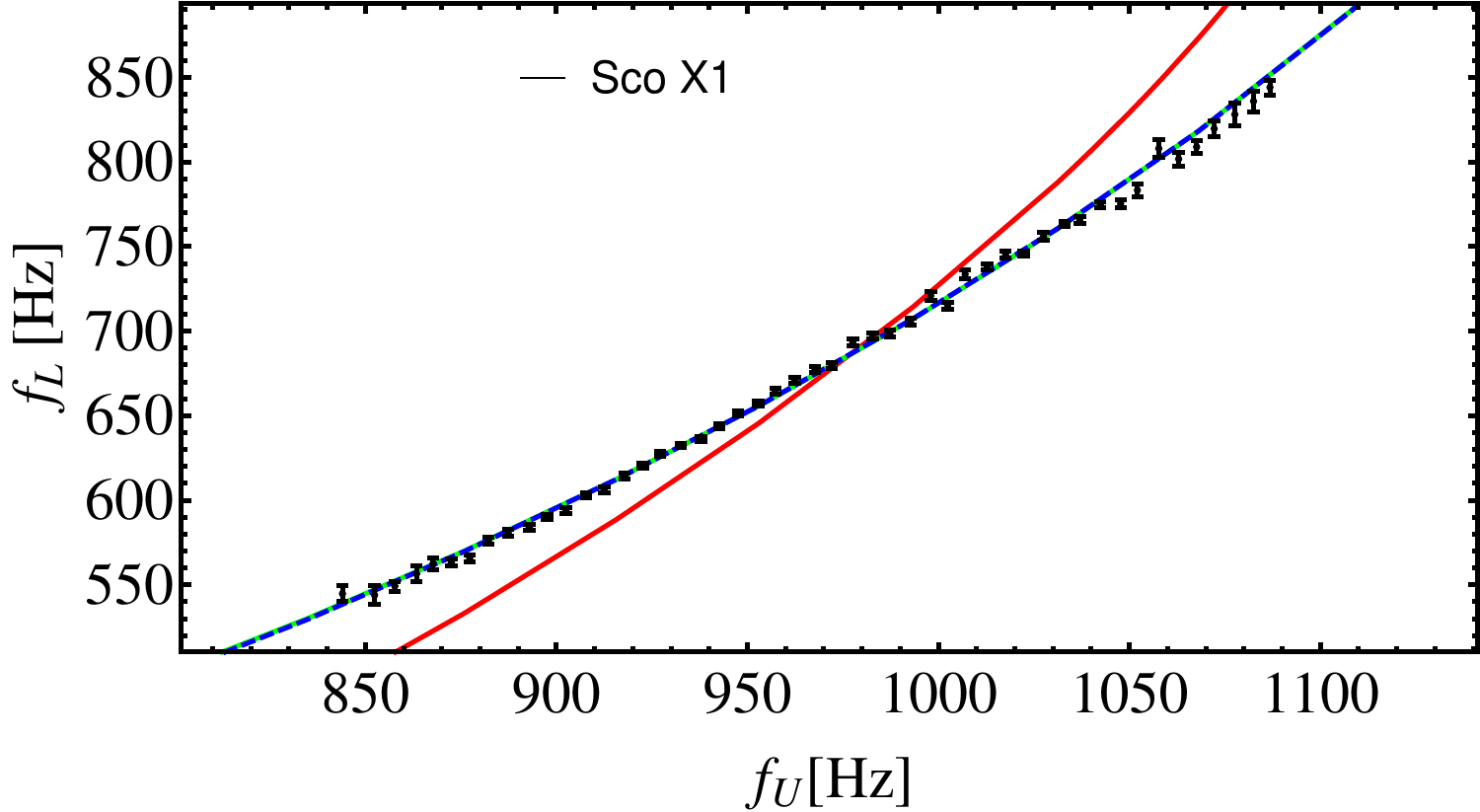}\hfill
\includegraphics[width=0.42\hsize,clip]{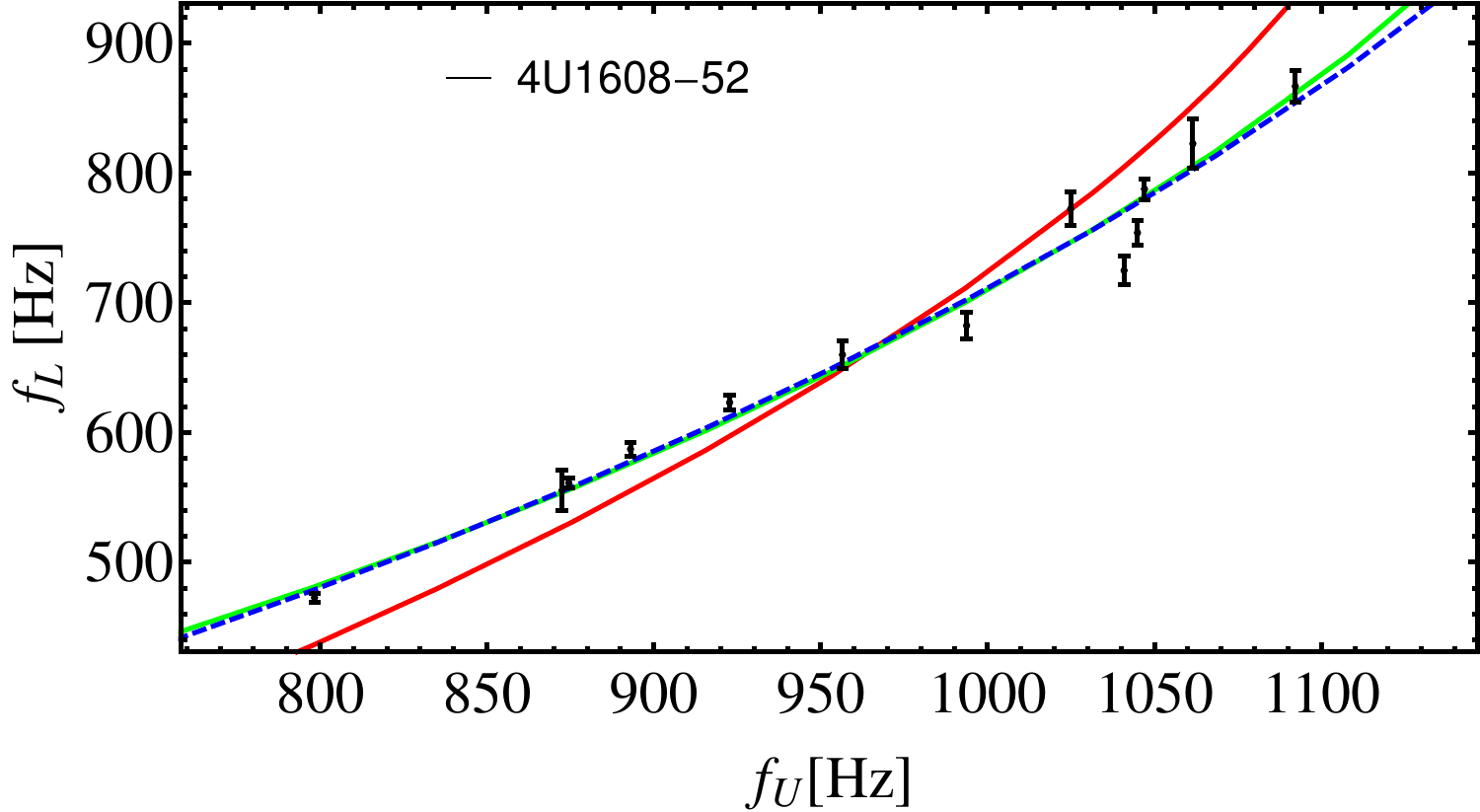}\hfill}

{\hfill
\includegraphics[width=0.42\hsize,clip]{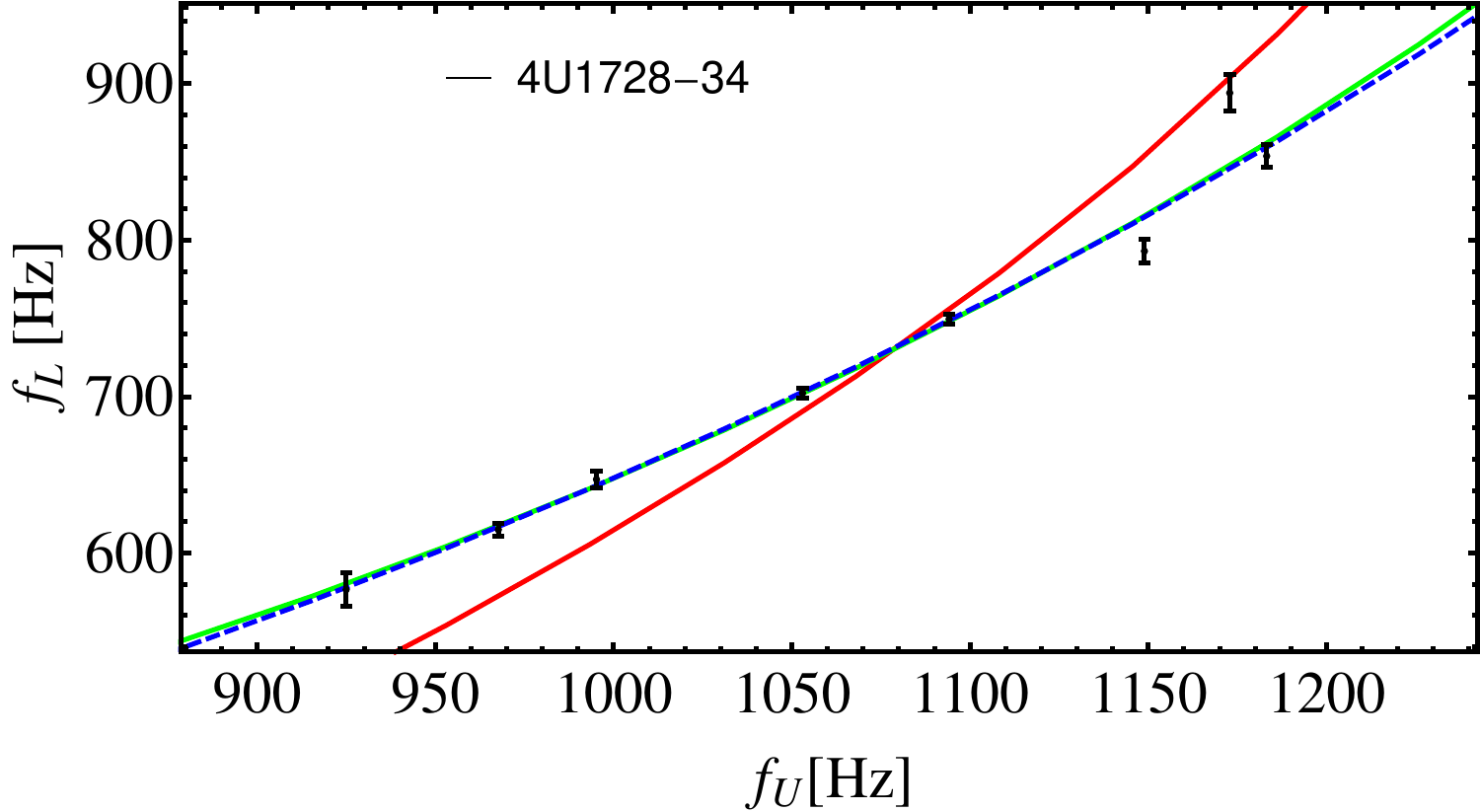}\hfill
\includegraphics[width=0.42\hsize,clip]{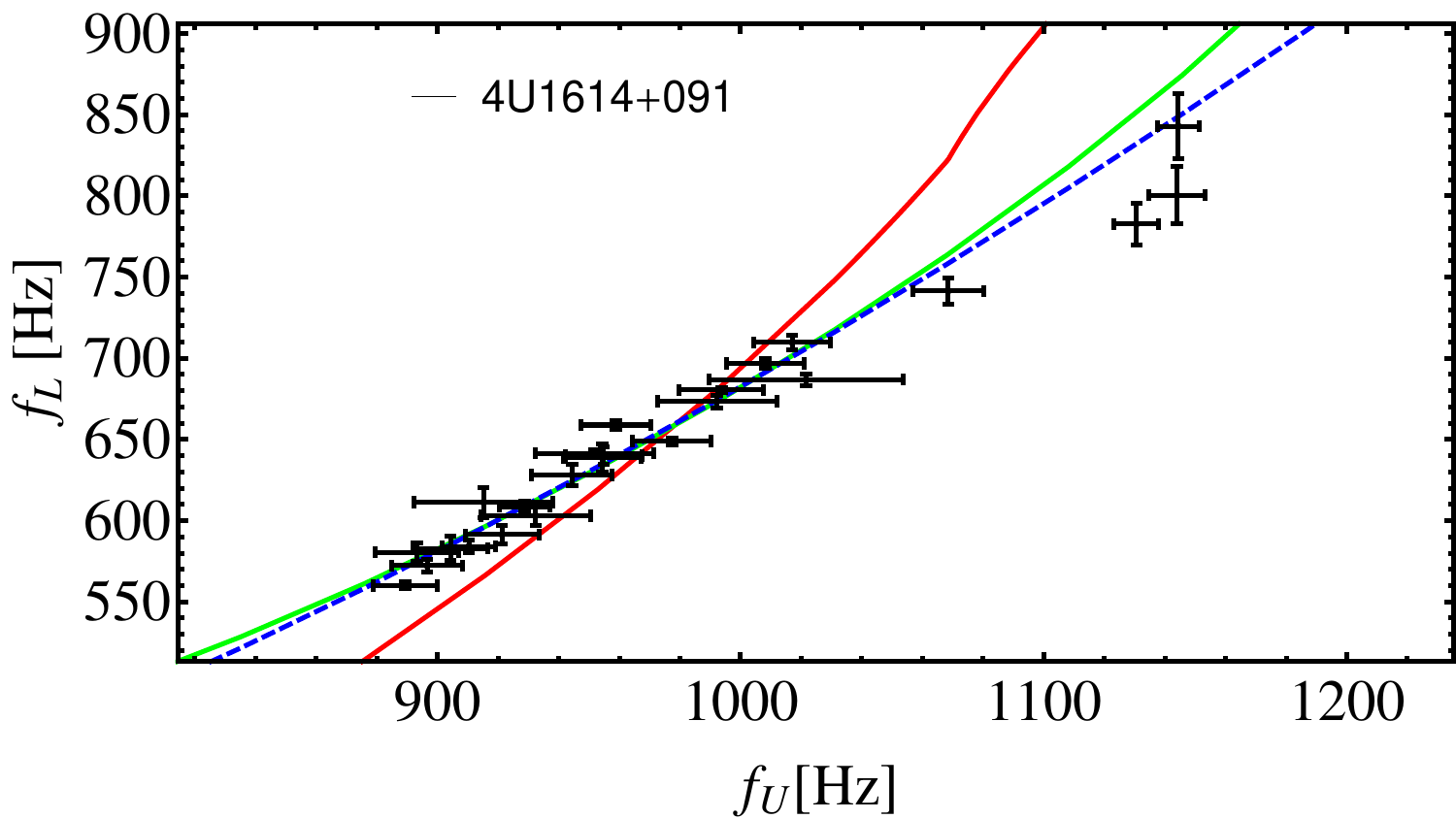}\hfill}
\caption{Fits of the lower frequencies $f_{\rm L}$ vs the upper frequencies $f_{\rm U}$ from the QPO data sets of the sources considered in this work (black data with error bars). Fits have been performed by considering Schwarzschild (red curve), Schwarzschild--de Sitter or anti-de Sitter (green curve) and Reisser-Nordstr\"{o}m (dashed blue curve) spacetimes.}
\label{fig:freq}
\end{figure*}

The results, summarized in Tab.~\ref{tab:results}, are statistically, observationally and theoretically analyzed below, for each source. The contour plots of the best-suited model parameters are shown in Fig.~\ref{fig:contours}.
For each source, the fits of the QPO lower $f_{\rm L}$ and upper $f_{\rm U}$ frequencies for the three models considered in this work are shown in Fig.~\ref{fig:freq}.

\begin{itemize}
\item[-] {\bf Cir~X-1}. Model 3 is strongly preferred over the other ones. However, from model 3 we get only a lower limit on the mass, \textit{i.e.}, $M>3$~M$_\odot$, which is incompatible with the NS interpretation supported by Ref.~\cite{2010ApJ...714..748T}. Model 2, instead, performs better than model 1 and provides a well-constrained mass up to 2-$\sigma$. In view of these considerations, model 2 is considered the best one. The data points for this source have been taken from Ref.~ \cite{2006ApJ...653.1435B}
\item[-] {\bf GX~5-1} \cite{1998ApJ...504L..35W,2002MNRAS.333..665J}, {\bf GX~17+2} \cite{2002ApJ...568..878H}, and {\bf GX~340+0} \cite{2000ApJ...537..374J}. For these sources, models 2 and 3 provide good fits to the data, with model 2 weakly favored with respect to model 3. However, in all the cases, model 3 does not constrain at all $\mathcal{C}$, being compatible with zero within 1-$\sigma$. The only difference is a quite unexpected large value (in modulus) of $R_0$ for GX 17+2, while error bars and the corresponding relative errors are all similar. Therefore, being $\mathcal{C}\approx0$, model 2 can be considered the best fit for these sources.
\item[-] {\bf Sco~X1}. Models 2 and 3 provide good fits to the data of this source, with model 3 weakly favored with respect to model 2. This sources shows intriguing results, since it seems not to exclude \emph{a priori} either a net topological charge, or possible deviations from GR due to a large value of $R_0$ for both models 2 and 3. However, model 3 provides a mass much higher than the range $1.40$--$1.52$~M$_\odot$ obtained by modelling optical light curves of Sco~X1 \cite{2021MNRAS.508.1389C}, whereas model 2 mass is closer to the above observational range. Therefore, we conclude that model 2 best-fits the data of this source. The data points for this source have been taken from Ref.~ \cite{2000MNRAS.318..938M}.
\item[-] {\bf 4U1608-52}. In this case, models 3 is mildly preferred over model 2 and provides a NS mass which is considerably higher than (and inconsistent with) $2.07^{+0.25}_{-0.15}$~M$_\odot$, recently got from QPO data \cite{2020ApJ...899..139M}. On the other hand, within 2-$\sigma$, the charge parameters $\mathcal{C}$ is consistent with zero and the mass is consistent with the estimate from model 2, which is closer to the above recent finding from QPO \cite{2020ApJ...899..139M}. These considerations lead us to consider model 2 as the best-suited to describe the data of 4U1608-52. The data points for this source have been taken from Ref.~ \cite{1998ApJ...505L..23M}.
\item[-] {\bf 4U1728-34}. Model 2 is only weakly preferred over model 3. However, model 3 does not constrain at all $\mathcal{C}$, being compatible with zero within 1-$\sigma$ and, therefore, we conclude that model 2 is the best fit of the source. The data points for this source have been taken from Ref.~ \cite{1999ApJ...517L..51M}.
\item[-] {\bf 4U0614+091}. Model 3 is strongly preferred over the other ones, but provides a lower limit on the source mass of $3.5$~M$_\odot$, which is incompatible with the NS interpretation. This value is also incompatible with the above-mentioned finding from QPO \cite{2020ApJ...899..139M}. For this reason, we conclude that a suitable fit for the data is provided by model 2. The data points for this source have been taken from Ref.~ \cite{1997ApJ...486L..47F}.
\end{itemize}

Thus, in view of the above, we can summarize our findings as follows.

\begin{itemize}
    \item[-] The best metric is either Schwarzschild--de Sitter or anti-de Sitter for all the sources.
    \item[-] The Reissner-Nordstr\"{o}m term is always (statistically, observationally, or theoretically) disfavored indicating either no charge contribution or the absence of  departures from extension of gravity.
\end{itemize}

In particular, our results may suggest the presence of:

\begin{itemize}
    \item[-] no topological or standard electric charge, showing a global neutrality of NS, \textit{i.e.}, although the local neutrality is not excluded, it seems that globally the compact object does not exhibit effective charge;
    \item[-] an effective constant energy term, acting as a cosmological constant, but with huge orders of difference. This may be interpreted as a constant energy contribution to the net energy of the NS that modifies its stress-energy momentum and is necessary to describe {\it in toto} the NS.
\end{itemize}

For the sake of completeness, we clearly cannot avoid to discuss about possible drawbacks associated with the statistical analyses that we performed. In particular, the results appear  weak and/or biased by caveats for the following reasons.
\begin{itemize}
    \item[-] More sources are needed to confirm or disfavor the above results. The classes of sources, here-involved, exhibit different results among them. If the mass values seem to be almost stable and $\lesssim 3\,M_\odot$, the other terms appear very different among class of source, but similar within the same class.
    \item[-] The contours in Fig.~\ref{fig:contours} appear extremely tight at least for 4 sources out of 8. The plots in Fig.~\ref{fig:freq} do not provide fully-viable outcomes for all sources, likely as a consequence of the low number of data points present in the corresponding catalogs. This can be responsible for the strange occurrence for that we cannot conclude that either charge or GR extensions are not present. Refined analyses require more data to improve the error bars got from the above contours.
    \item[-] The most general involved metric fully-degenerates with standard cases of GR. In fact, it is possible to re-obtain the same results of fourth order $F(R)$ theories in GR when the curvature is constant, suggesting that the most suitable benchmark seems to be   GR.
    \item[-] Models 2, applied to each classes of data points, indicate that there is no consensus on the $\Lambda$ sign. Even though this appears licit in $F(R)$, in GR fixes the sign of vacuum energy, \textit{i.e.}, modifies the energy momentum tensor. Alternatively speaking, the NS sources seem to require a non-zero contribution to energy, but whose sign corresponds to either de Sitter and anti-de Sitter solutions.
\end{itemize}

\section{Conclusions and perspectives}\label{sezione5}

In this paper, we considered eight NS sources and we fitted the corresponding frequency data with three QPO models. In particular, we employed a spherically symmetric static spacetime constructed as a Reisser-Nordstr\"{o}m solution with the addition of the a de Sitter or anti-de Sitter contribution. The choice of the metric is discussed in view of both GR and fourth order extended theories of gravity. Further, we debated the degeneracy between the two frameworks and we stressed that the spacetime describes NS with a net external charge and a constant term of energy that contributes to the energy-momentum tensor.

The statistical analyses were performed through a MCMC code, based on the Metropolis-Hastings algorithm, that provided best-fit parameters and 1--$\sigma$ and 2--$\sigma$ contour plots and error bars. We demonstrated, statistically and theoretically, that the best suite for our results is always represented by the Schwarzschild de Sitter and anti-de Sitter solutions, though weak evidences in favor of the Reisser-Nordstr\"{o}m solution were found.

In view of the degeneracy between GR and fourth order gravity, we show that all the sources, here considered, fully agree with GR predictions and seem to exclude extended versions of  gravity. Numerical analyses, however, did not completely exclude to get departures from GR as the curvature is not assumed to be approximately constant, namely $R=R_0$. Theoretical discussions have been carried forward in view of the fact that there is no consensus on the sign of $R_0$, leading to de Sitter and anti-de Sitter phases, depending on the particular source under analysis. Moreover, possible drawbacks of our approach have been also discussed, \textit{i.e.}, the need of more data points and source.

Further considerations will be investigated in future efforts. In particular, in view of the small number of data points for each source, we can perform mock compilations of data adopting more refined techniques, e.g., machine learning. Alternative spacetimes from modified theories of gravity can be also tested, showing whether they could agree with these data and improve the standard fits performed using the Schwarzschild solution. Finally, we will investigate the effects of non-singular metrics on the overall analysis, checking whether they can improve the quality of our findings.

\section*{Acknowledgements}
OL  is grateful to the Department of Physics of the Al-Farabi University for hospitality during the period in which this manuscript has been written. OL acknowledges Roberto Giambò for fruitful discussions on the subject of this paper. KB acknowledges Mariano Mendez for providing QPO data. This research has been partially funded by the Science Committee of the Ministry of Science and Higher Education of the Republic of Kazakhstan (Grant No. AP19680128).

\end{document}